\documentclass[%
 reprint,
 amsmath,amssymb,
 aps,
]{revtex4-2}

\usepackage{graphicx}
\usepackage{dcolumn}
\usepackage{bm}
\usepackage{hyperref}
\usepackage{slashed}
\usepackage{subfloat}
\usepackage{pgfplots}
\usepackage{subfigure}
\usepackage{etoolbox}   
\usepackage{geometry}
\usepackage{booktabs}
\definecolor{acsblue}{RGB}{17,76,139}

\begin{document}

\fontsize{7.6}{8.6}\selectfont


\title{Non-Hermitian $\mathcal{PT}$ symmetric Klein-Gordon fields in Lorentz-violating wormholes spacetime  background}

\author{Omar Mustafa
}
\email{omar.mustafa@emu.edu.tr  (Corresponding Author)}
\affiliation{Department of Physics, Eastern Mediterranean University, 99628, G. Magusa, north Cyprus, Mersin 10 - Türkiye}

\author{Abdullah Guvendi
}
\email{abdullah.guvendi@erzurum.edu.tr }
\affiliation{Department of Basic Sciences, Erzurum Technical University, 25050, Erzurum, Türkiye}


\begin{abstract}
{\fontsize{7.6}{8.6}\selectfont \setlength{\parindent}{0pt}
We investigate a $\mathcal{PT}$-symmetric Klein--Gordon (KG) oscillator in a $(3+1)$-dimensional Lorentz-violating (LV) traversable wormhole spacetime. The geometry is characterized by a smooth throat of radius $a$ and a constant lapse function, $\mathcal{A}(x)=1$, so that no Killing horizons occur and the two asymptotic regions at \(x\to \pm\infty\) remain connected through the throat. The Lorentz violation modifies the radial geometry and the oscillator scale according to $\widetilde{\Omega}=\Omega/\sqrt{1-\zeta}$. A nonminimal non-Hermitian coupling, $\mathcal{F}_t(x)=i\widetilde{\Omega}x$, leads to a $\mathcal{PT}$-symmetric KG oscillator with a regular effective potential at the throat and a quadratic confinement at large $|x|$. The radial equation can be mapped to the confluent Heun equation, and polynomial truncation gives a conditionally exactly solvable sector in which the oscillator frequency, throat radius, LV parameter, angular momentum, and radial quantum number satisfy algebraic constraints. For the degree-1 sector, these conditions restrict the allowed angular momenta and correlate them with the LV parameter. The resulting bound-state energies are real and discrete, with exact particle--antiparticle symmetry, $E^{-}=-E^{+}$, while the corresponding wave functions are regular and localized on the two-sided wormhole geometry. These results show how Lorentz violation and the $\mathcal{PT}$-symmetric coupling modify the relativistic bound-state spectrum through the geometry of the wormhole.}
\end{abstract}

\keywords{Lorentz-violating wormholes; $\mathcal{PT}$ symmetry; Klein--Gordon oscillators; confluent Heun equations; conditional exact solvability; relativistic bound states}

\maketitle


\section{Introduction}
\label{sec:1}

\setlength{\parindent}{0pt}

Lorentz invariance is a fundamental symmetry of the Standard Model and General Relativity, determining the local causal structure of spacetime and the invariant propagation speed of light \cite{Kibble,Smolin,C-1,C-2}. Nevertheless, departures from exact Lorentz symmetry arise in several approaches to quantum gravity and extensions of high-energy field theory \cite{Kibble,Smolin,C-1,C-2,C-3,C-4,C-5}. The Standard-Model Extension (SME) provides a general effective-field-theory framework for parametrizing Lorentz and charge-parity-time (CPT) violation through background tensor fields that select preferred spacetime directions while preserving observer covariance \cite{C-1,C-2,KS1989,ColladayKostelecky1997}. Precision experiments and theoretical analyses have consequently placed stringent bounds on the corresponding coefficients across a broad range of sectors, including CPT tests \cite{23,24,25,26}, radiative corrections \cite{27,28,29,30,31,32,33,34}, gauge-field dynamics \cite{35,36,37,38,39}, photon-fermion interactions \cite{40,41,42,43}, fermionic couplings \cite{44,45,46,47}, and the quantum information content of spacetime \cite{SUGG}.

\setlength{\parindent}{0pt}
\vspace{1pt}

Modified gravitational theories provide a further setting in which Lorentz violation can affect spacetime structure. In particular, nontrivial topologies can arise without requiring conventional matter that violates the classical energy conditions, allowing traversable wormhole geometries \cite{TW1,TW2,TW3,TW4,Morris,TW5,Falco1,Falco2,Falco3}. In such theories, the effective energy-condition violation can originate from modified geometric terms, including higher-curvature invariants, nonminimal couplings, or background tensor fields \cite{TW6,TW7,TW8,TW9,Radhakrishnan2024}. Einstein-bumblebee gravity provides a representative realization \cite{bumblebee,Eslam,Ovgun2019,Oliveira2018,Ding2024,Maluf2020,Ghosh2024,MalufMuniz2021,KS1989,C-1,LV}, in which a dynamical vector field acquires a nonzero vacuum expectation value and induces spontaneous Lorentz violation. Its field equations admit traversable wormhole solutions whose throat geometry depends explicitly on the LV parameter \cite{Ovgun2019}. Analytically tractable treatments of quantum fields in curved spacetime further provide a framework for studying curvature-induced effective potentials \cite{Guvendi-PLB-1,Guvendi-PLB-2,abdelghani-2025,o1}, bound states, and topological effects. In LV wormhole geometries \cite{LV,LV-1,LV-2}, such curvature-induced potentials can substantially modify vector-boson modes \cite{LV-2}, leading to discrete spectra and modified vacuum fluctuations near the throat.

\setlength{\parindent}{0pt}

\begin{figure*}[t]
\centering
\includegraphics[ width=0.78\textwidth ]{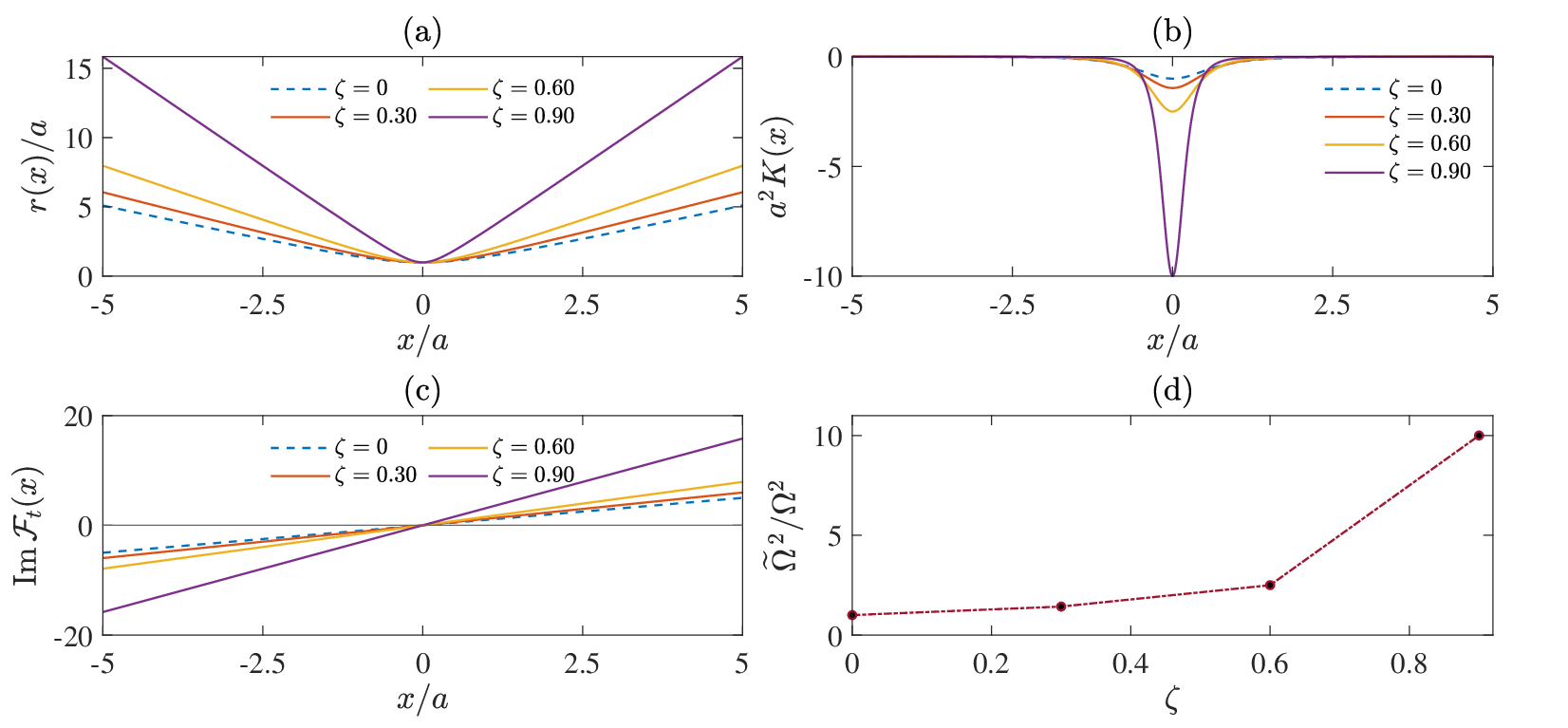}
\caption{\fontsize{7.6}{8.6}\selectfont \textbf{LV wormhole geometry and $\mathcal{PT}$-symmetric oscillator structure.} (a) Dimensionless areal radius $r(x)/a$, with $r(x)=\sqrt{a^2+x^2/(1-\zeta)}$. (b) Dimensionless Gaussian curvature $a^2K(x)$, where $K(x)=-a^2(1-\zeta)/[a^2(1-\zeta)+x^2]^2$ \cite{LV-1,LV-2}. (c) Dimensionless imaginary component of the non-Hermitian coupling $\mathcal{F}_t(x)=i\widetilde{\Omega}x$, with $\widetilde{\Omega} =\Omega/\sqrt{1-\zeta}$; its odd spatial dependence is consistent with $\mathcal{F}_t^{\ast}(-x)=\mathcal{F}_t(x)$. (d) Lorentz-violation enhancement of the oscillator scale,$\widetilde{\Omega}^{\,2}/\Omega^2=(1-\zeta)^{-1}$.The curves correspond to $\zeta=0$, $0.30$, $0.60$, and $0.90$, with $a=1$, $\Omega=1$, $x/a\in[-5,5]$, and $0\leq\zeta<1$ in the adopted dimensionless units.} \label{fig:introduction}
\end{figure*}

\setlength{\parindent}{0pt}
\vspace{1pt}

Here we study spin-$0$ scalar bosonic oscillator fields in a four-dimensional traversable LV wormhole described by
\begin{equation}
ds^2=-\mathcal{A}(x)\,dt^2+\frac{1}{\mathcal{A}(x)}\,dx^2+r(x)^2\left(d\theta^2+\sin^2\theta\,d\varphi^2
\right),\label{1.1}
\end{equation}
where $\mathcal{A}(x)$ is the redshift function and $r(x)$ is the areal radius \cite{Guvendi-PLB-1,LV}. We determine how this background modifies the radial modes and spectrum of scalar particles and antiparticles in a non-Hermitian $\mathcal{PT}$-symmetric configuration. The analysis focuses on the combined effects of curvature, Lorentz violation, and the non-Hermitian oscillator coupling on the relativistic spectrum.

\setlength{\parindent}{0pt}
\vspace{1pt}

Non-Hermitian quantum mechanics was placed on a physical footing by the demonstration that Hamiltonians invariant under combined parity and time reversal can possess real spectra despite being non-Hermitian \cite{PT1}. Since the early experimental studies, $\mathcal{PT}$-symmetric systems have been realized in optical waveguides, lasers, optical resonators, microwave cavities, superconducting circuits, nuclear magnetic resonance, graphene, and metamaterials \cite{PT11}. In the unbroken phase, spectral reality is associated with $\mathcal{PT}$ symmetry rather than the Hermiticity condition $H=H^\dagger$. For a $\mathcal{PT}$-symmetric Hamiltonian,
\begin{equation}
H=H^\ddagger=\mathcal{PT}H\mathcal{PT},
\end{equation}
with parity and time reversal acting as
\begin{equation}
\mathcal{P}x\mathcal{P}=-x,
\qquad
\mathcal{T}i\mathcal{T}=-i,
\end{equation}
respectively \cite{PT1,PT2,PT3}. Reviews by El-Ganainy et al. \cite{PT4} and Bender \cite{PT11} summarize the development and experimental relevance of this framework.

We introduce a $\mathcal{PT}$-symmetric nonminimal scalar coupling
\begin{equation}
\mathcal{F}_t(x)
=
i\widetilde{\Omega}x,
\qquad
\widetilde{\Omega}
=
\frac{\Omega}{\sqrt{1-\zeta}},
\end{equation}
where $\zeta$ is the Lorentz-violation parameter. It satisfies
\begin{equation}
\mathcal{F}_t^{\ast}(-x)
=
\mathcal{F}_t(x),
\end{equation}
and therefore defines an exactly $\mathcal{PT}$-symmetric interaction sector. The same parameter $\zeta$ that deforms the wormhole geometry also controls the effective oscillator scale, with
\[
\frac{\widetilde{\Omega}^{\,2}}{\Omega^2}
=
\frac{1}{1-\zeta},
\]
as shown in Fig.~\ref{fig:introduction}. To the best of our knowledge, this provides a previously unexplored realization of a non-Hermitian $\mathcal{PT}$-symmetric scalar-bosonic oscillator coupled to a LV traversable-wormhole geometry.

\setlength{\parindent}{0pt}
\vspace{1pt}

Fig.~\ref{fig:introduction} summarizes the geometric and coupling structure underlying the model. The finite throat remains fixed at
$r(0)=a$, whereas increasing $\zeta$ changes the radial profile and the curvature distribution. At the same time, $\widetilde{\Omega}$ increases monotonically with $\zeta$, strengthening the oscillator coupling while preserving the exact condition
$\mathcal{F}_t^{\ast}(-x)=\mathcal{F}_t(x)$. Thus, the LV parameter provides a single control parameter linking the wormhole geometry to the non-Hermitian oscillator scale.

\setlength{\parindent}{0pt}
\vspace{1pt}

The paper is organized as follows. Section~\ref{sec:2} develops the KG dynamics in the spacetime (\ref{1.1}), with $\mathcal{A}(x)=1$. This choice removes Killing horizons and gives a globally traversable geometry. We consider the nonminimal four-vector interaction
\[\mathcal{F}_\mu=\bigl(\mathcal{F}_t(x),0,0,0\bigr)\]
\cite{barbosa2025,barbosa2026}, with
$\mathcal{F}_t(x)=i\widetilde{\Omega}x$, and show that it generates an effective $\mathcal{PT}$-symmetric KG oscillator. The resulting potential remains regular at the throat, where the finite wormhole radius replaces the singular centrifugal behaviour associated with the flat-space origin. Section~\ref{sec:3} reduces the radial equation to a confluent Heun system and obtains conditionally exact solutions in terms of confluent Heun polynomials \cite{Heun1}, subject to the required parameter correlations \cite{Levai}. The resulting solvable sector possesses a real discrete spectrum subject to the algebraic constraints imposed by the confluent-Heun polynomial conditions, with its levels controlled by the curvature, Lorentz-violation parameter, and oscillator coupling. Section~\ref{sec:4} summarizes the main results and their physical implications.

\section{$(\mathcal{PT})$-symmetric scalar bosonic field
$(\mathcal{F}_t(x)=i\tilde{\Omega}x)$ in a LV wormhole}
\label{sec:2}

\setlength{\parindent}{0pt}

Wormhole geometries provide a well-defined setting in which quantum-field
dynamics can be studied on curved backgrounds with a nontrivial global
topology. LV deformations further enlarge this class of
geometries by modifying the radial structure while retaining sufficient
symmetry for an analytic treatment \cite{Guvendi-PLB-1,LV}. We therefore
consider a scalar bosonic oscillator field propagating in a static,
spherically symmetric LV wormhole described by the line element
$ds^2=g_{\mu\nu}dx^\mu dx^\nu$ in Eq.~(\ref{1.1}). The metric tensor has the
diagonal form
\begin{equation}
g_{\mu\nu}=\mathrm{diag}\left(-\mathcal{A}(x),\mathcal{A}(x)^{-1},r(x)^2,r(x)^2\sin^2\theta\right),
\end{equation}
with inverse metric
\begin{equation}
g^{\mu\nu}=\mathrm{diag}\left(-\mathcal{A}(x)^{-1},\mathcal{A}(x),r(x)^{-2},r(x)^{-2}\sin^{-2}\theta\right).
\end{equation}
The coordinates $(t,x,\theta,\varphi)$ cover the two-sided radial manifold,
with $t,x\in(-\infty,\infty)$, $\theta\in(0,\pi)$, and
$\varphi\in(0,2\pi)$. The metric determinant is
\begin{equation}
\det(g_{\mu\nu})=g=-r(x)^4\sin^2\theta,\qquad\sqrt{-g}=r(x)^2\sin\theta.
\end{equation}
The areal radius $r(x)$ determines the radial geometry and, together with the
redshift function $\mathcal{A}(x)$, specifies the geometric background seen
by the scalar field. At the minimal-radius hypersurface ($x=0$), the throat conditions ($r(0)=a$), ($r'(0)=0$), and ($r''(0)>0$) identify a smooth, non-degenerate throat and permit a reflection-symmetric extension of the radial coordinate through ($x=0$). Thus, ($a>0$) fixes the minimum areal radius and distinguishes the wormhole geometry from a conventional radial coordinate origin. For the LV solution considered here, the redshift function is constrained by \(\mathcal{A}''(x)=0\), which gives the affine family
\begin{equation}
\mathcal{A}(x)=a_1+a_2x.
\end{equation}
We restrict attention to the constant-lapse configuration $a_1=1$ and $a_2=0$, for which
\begin{equation}
\mathcal{A}(x)=1.
\end{equation}
This choice removes Killing horizons and yields a globally static geometry whose two asymptotic regions are connected through the regular throat. The areal radius is then taken to \cite{Guvendi-PLB-1,LV}
\begin{equation}
r(x)
=
\sqrt{
a^2+\frac{x^2}{1-\zeta}
},
\qquad
0\leq\zeta<1.
\label{2.01}
\end{equation}
The parameter $\zeta$ therefore controls the radial stretching away from the throat while leaving the throat radius itself fixed at $r(0)=a$.. The restriction $0\leq\zeta<1$ ensures that the radial deformation and the effective parameters introduced below remain real and finite. In particular, $r(0)=a$ and $r(x)=r(-x)$, whereas the two asymptotic regions are reached as $x\rightarrow\pm\infty$. Together with $a>0$ and $\mathcal{A}(x)=1$, this choice provides the regular traversable background in which the scalar KG modes are studied.

\begin{figure*}[ht!]
\centering
\includegraphics[width=0.90\textwidth]{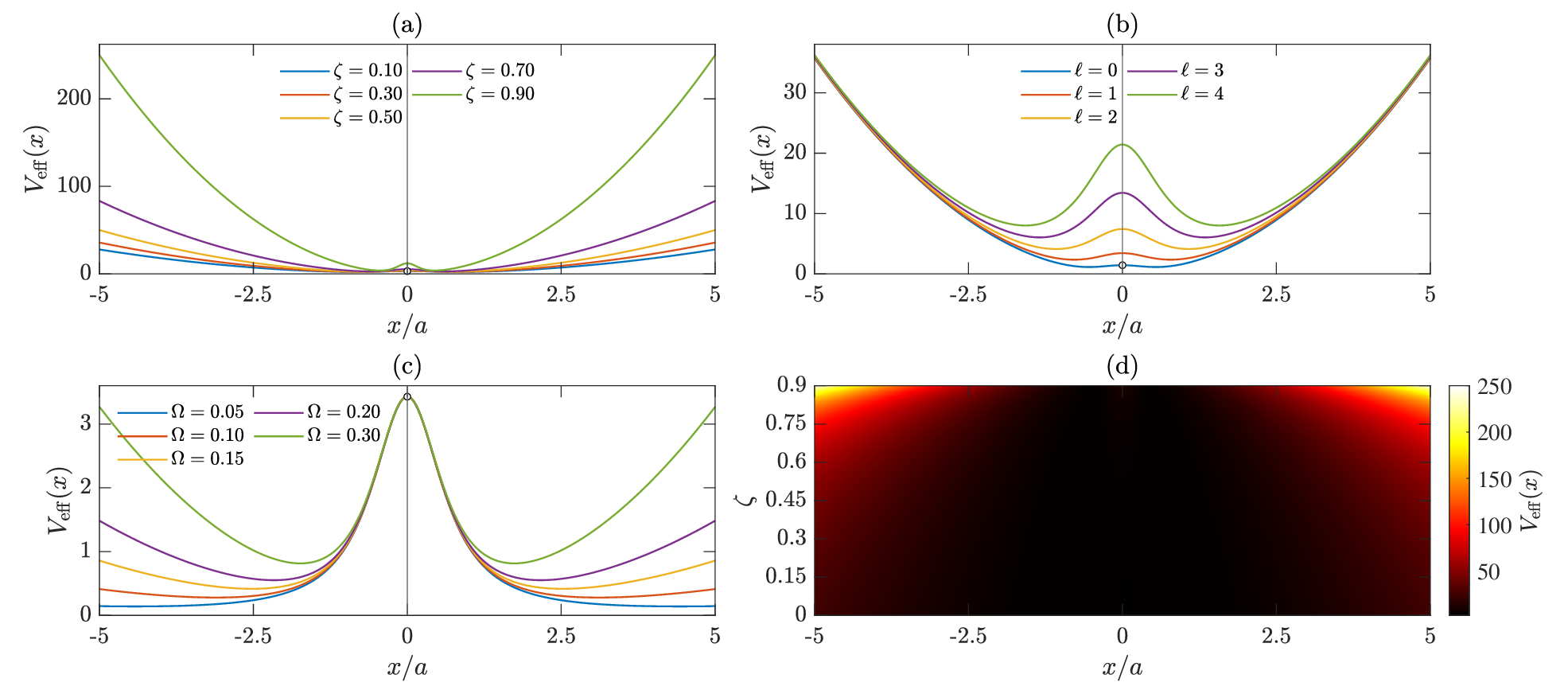}
\caption{\fontsize{7.6}{8.6}\selectfont
\setlength{\parindent}{0pt}
Effective potential $V_{\mathrm{eff}}(x)$ in Eq.~(\ref{eff}) for the
relativistic KG oscillator in the LV traversable wormhole
background. The potential depends on the Lorentz-violation parameter
$\zeta$, angular momentum quantum number $\ell$, oscillator frequency
$\Omega$, and throat parameter $a$. In all panels, $a=1$. (a)
$\zeta=0.1,0.3,0.5,0.7,0.9$ for $\ell=1$ and $\Omega=1$. (b)
$\ell=0,1,2,3,4$ for $\zeta=0.3$ and $\Omega=1$. (c)
$\Omega=0.05,0.10,0.15,0.20,0.30$ for $\zeta=0.3$ and $\ell=1$. (d) $\ell=1$ and $\Omega=1$.
For $0\leq\zeta<1$ and $a>0$, the potential is finite at the throat and
grows quadratically for large $|x|$, defining a regular confining radial
problem.}
\label{fig:1}
\end{figure*}

Within this background, the scalar bosonic field obeys the generalized
KG equation
\begin{equation}
\frac{1}{\sqrt{-g}}
\left(
\mathcal{D}^{+}_{\mu}
\sqrt{-g}\,
g^{\mu\nu}
\mathcal{D}^{-}_{\nu}
\right)
\Psi
=
m_0^2\Psi,
\label{2.2}
\end{equation}
where
\begin{equation}
\mathcal{D}^{\pm}_{\mu}
=
\partial_{\mu}\pm\mathcal{F}_{\mu}
\end{equation}
incorporate the nonminimal four-vector coupling
\begin{equation}
\mathcal{F}_{\mu}
=
\left(
\mathcal{F}_{t}(x),0,0,0
\right)
\end{equation}
\cite{barbosa2025,barbosa2026}. The temporal component is therefore the
only nonvanishing component of the nonminimal interaction. The
non-Hermitian structure enters through the temporal sector, while the
spatial derivatives retain their standard geometric form.

Using the separation ansatz
\begin{equation}
\Psi
=
e^{-iEt}
\mathcal{R}(x)
\mathcal{Y}_{\ell m}(\theta,\varphi),
\end{equation}
the field equation reduces to the radial equation
\begin{equation}
\left(
E^2-m_0^2+\mathcal{F}_t(x)^2
+\frac{1}{r(x)^2}
\partial_x r(x)^2\partial_x
-\frac{\ell(\ell+1)}{r(x)^2}
\right)
\mathcal{R}(x)
=
0.
\label{2.3}
\end{equation}

We now impose the non-Hermitian
$\mathcal{PT}$-symmetric coupling
\begin{equation}
\mathcal{F}_t(x)
=
i\tilde{\Omega}x,
\end{equation}
so that
$\mathcal{D}^{\pm}_t=\partial_t\pm i\tilde{\Omega}x$.
The resulting radial equation takes the form
\begin{equation}
\begin{split}
\mathcal{R}''(x)
+
\frac{2x}{x^2+\tilde a^2}\mathcal{R}'(x)
+
\left[
\mathcal{E}
-\tilde{\Omega}^2x^2
-\frac{\tilde{\ell}(\tilde{\ell}+1)}
{x^2+\tilde a^2}
\right]
\mathcal{R}(x)
=
0,
\end{split}
\label{2.4}
\end{equation}
where
\begin{equation}
\begin{split}
&
\mathcal{E}=E^2-m_0^2,
\qquad
\tilde a^2=a^2(1-\zeta),
\qquad
\tilde{\Omega}^2=\frac{\Omega^2}{1-\zeta}.
\end{split}
\end{equation}
The effective angular parameter is defined through
\begin{equation}
\tilde{\ell}(\tilde{\ell}+1)
=
(1-\zeta)\ell(\ell+1)
\Rightarrow
\tilde{\ell}
=
-\frac{1}{2}
+
\sqrt{
\frac{1}{4}
+
(1-\zeta)\ell(\ell+1)
}
\geq0.
\end{equation}
Thus, the LV deformation enters the radial dynamics through the effective throat scale $\tilde a$, the oscillator scale $\tilde{\Omega}$, and the angular contribution encoded by $\tilde{\ell}$. The limiting structure provides a direct consistency check on the construction. When $a\rightarrow0$ and $\zeta\rightarrow0$, Eq.~(\ref{2.01}) gives $r(x)\rightarrow |x|$; restricting to the conventional radial half-line $x>0$ then yields $r(x)\rightarrow x$. Simultaneously,
$\tilde a\rightarrow0$,
$\tilde{\Omega}\rightarrow\Omega$, and
$\tilde{\ell}\rightarrow\ell$.
Equation~(\ref{2.4}) consequently approaches the standard flat-space
radial KG oscillator equation. The present system therefore
constitutes a geometric and LV deformation of the familiar
radial oscillator rather than an unrelated dynamical construction. The
finite-$a$ wormhole case, however, is defined on the complete line
$x\in(-\infty,\infty)$, so that the two asymptotic regions remain connected
through the regular throat.

To expose the spectral structure, Eq.~(\ref{2.4}) is transformed to a
one-dimensional Schr\"odinger-type equation by
\begin{equation}
\mathcal{R}(x)
=
\frac{U(x)}
{\sqrt{\tilde a^2+x^2}},
\end{equation}
which gives
\begin{equation}
U''(x)
+
\left[
\mathcal{E}
-
V_{\mathrm{eff}}(x)
\right]
U(x)
=
0,
\end{equation}
with
\begin{equation}
V_{\mathrm{eff}}(x)
=
\tilde{\Omega}^2x^2
+
\frac{\tilde{\ell}(\tilde{\ell}+1)}
{\tilde a^2+x^2}
+
\frac{\tilde a^2}
{(\tilde a^2+x^2)^2}.
\label{eff}
\end{equation}

The effective potential is an even, real function of $x$ and hence satisfies
\begin{equation}
V_{\mathrm{eff}}(x)
=
V_{\mathrm{eff}}^{*}(-x),
\end{equation}
consistent with the $\mathcal{PT}$ symmetry of the radial problem. Because
$\tilde a>0$ for $a>0$ and $0\leq\zeta<1$, all terms in
Eq.~(\ref{eff}) remain finite at the throat. In particular, the value at
$x=0$ is
\begin{equation}
V_{\mathrm{eff}}(0)
=
\frac{\tilde{\ell}(\tilde{\ell}+1)}
{\tilde a^2}
+
\frac{1}{\tilde a^2},
\end{equation}
whereas
\begin{equation}
V_{\mathrm{eff}}(x)
\sim
\tilde{\Omega}^{\,2}x^2,
\qquad
|x|\rightarrow\infty.
\end{equation}
The radial problem is therefore regular at the throat and confining in both
asymptotic directions.

Fig.~\ref{fig:1} summarizes the parameter dependence of the resulting
effective potential. Panel (a) isolates the Lorentz-violation
dependence at fixed $\ell$ and $\Omega$. Increasing $\zeta$ decreases the
effective throat scale
$\tilde a^2=a^2(1-\zeta)$ while increasing the oscillator scale
$\tilde{\Omega}^2=\Omega^2/(1-\zeta)$. The LV deformation therefore changes
both the finite-throat contribution and the asymptotic confining term.
 Panel (b) displays the dependence on the angular moment. Since
$\tilde{\ell}(\tilde{\ell}+1)=(1-\zeta)\ell(\ell+1)$, increasing $\ell$
increases the positive angular contribution to potential while leaving
the leading quadratic asymptotic term unchanged at fixed $\Omega$ and
$\zeta$. The panel (c) isolates the oscillator frequency: increasing
$\Omega$ directly strengthens the quadratic contribution and consequently
increases the asymptotic confinement. Panel (d) displays the potential magnitude for different values of $\zeta$. These parameter dependencies follow
directly from Eq.~(\ref{eff}) and provide the relevant scales for the
spectral problem considered in the following.

The finite throat also changes the mathematical domain of the radial
problem. Unlike the ordinary flat-space radial equation, for which the
origin is associated with the endpoint of the half-line and the centrifugal
term may become singular, $x=0$ is an interior and regular point of the
wormhole manifold. Consequently, the admissible radial solutions are
selected by regularity together with the appropriate matching or parity
conditions across the throat, rather than by imposing a singular-origin
condition solely on geometric grounds. Eq.~(\ref{eff}) therefore
permits regular modes to extend continuously through $x=0$ and to connect
the two asymptotic regions. The throat modifies not only the local shape of
the effective potential, but also the domain and boundary structure of the
associated spectral problem.

The parameters $\zeta$, $a$, $\ell$, and $\Omega$ consequently enter the
radial dynamics through distinct and identifiable mechanisms. The
Lorentz-violation parameter deforms the radial geometry and simultaneously
rescales the effective oscillator strength; the throat parameter fixes the
minimum areal radius and removes the origin singularity; the angular
quantum number controls the geometric angular contribution; and the
oscillator frequency determines the strength of the asymptotic quadratic
confinement. Their combined dependence is encoded explicitly in
Eq.~(\ref{eff}), providing the starting point for the confluent-Heun
reduction and conditional quantization developed in
Sec.~\ref{sec:3}.

\section{Corresponding energies and wave functions}
\label{sec:3}

\setlength{\parindent}{0pt}

We proceed with the traditional oscillator analogy and use
$\mathcal{R}(x)=e^{-{\tilde\Omega x^2}/{2}}\mathcal{H}(x)$
in Eq.~(\ref{2.4}), followed by the change of variable
$y=-x^2/\tilde a^2 \Rightarrow x=\tilde a\sqrt{-y}$, to obtain
\begin{equation}
\begin{split}
&(y-1)\mathcal{H}''+
\left[\alpha y+P-\frac{(\beta+1)}{y}\right]\mathcal{H}'
+\left[(\mu+\nu)-\frac{\mu}{y}\right]\mathcal{H}=0,\\
&\mu+\nu=\frac{\tilde a^2}{4}\left(3\tilde\Omega-\mathcal E\right),
\qquad
\mu=\frac{1}{4}
\left[
\tilde\ell(\tilde\ell+1)
+\tilde a^2\left(\tilde\Omega-\mathcal E\right)
\right],\\
&\alpha=\tilde a^2\tilde\Omega,
\qquad
P=\frac{3}{2}-\tilde a^2\tilde\Omega,
\qquad
\beta=-\frac{1}{2}.
\end{split}
\label{2.5}
\end{equation}

\begin{figure*}[ht!]
\centering
\includegraphics[width=0.99\textwidth]{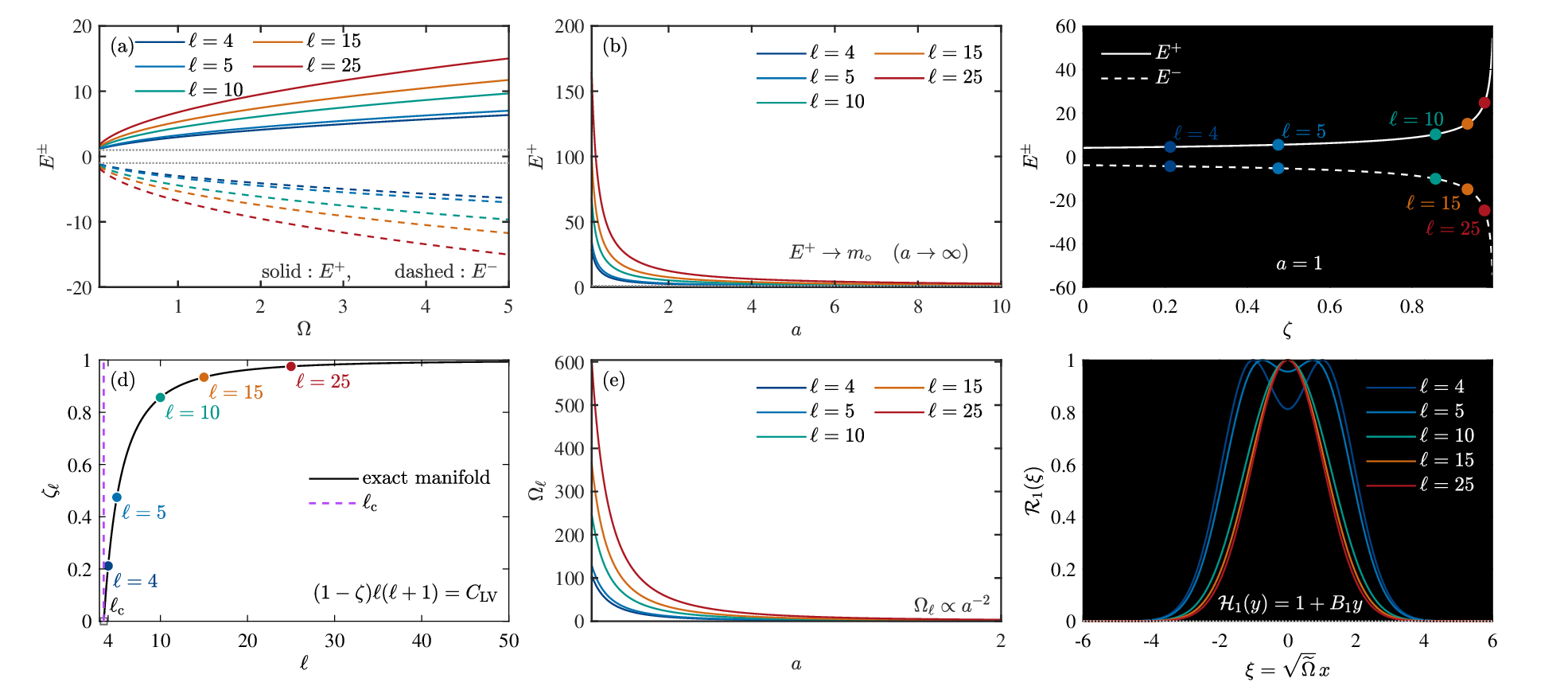}
\caption{\fontsize{7.6}{8.6}\selectfont \textbf{Conditionally exact spectrum and degree-1 Heun states.}
We set $m_{\circ}=1$, $n=0$, $\alpha_{\mathrm H}=(25-\sqrt{65})/8$, $B_{1}=-(\sqrt{65}+3)/4$, and $C_{\mathrm{LV}}=(55+\sqrt{65})/4$, with admissible $\ell={4,5,10,15,25}$ and $\zeta_{\ell}=1-C_{\mathrm{LV}}/[\ell(\ell+1)]$. (a) $E^{\pm}(\Omega)$ for $\Omega\in[0.05,5]$; (b) $E^{+}(a)$ for $a\in[0.15,10]$ under the exact constraint; (c) $E^{\pm}(\zeta)$ at $a=1$, with the allowed $\ell$ states marked; (d) conditional-solvability manifold $(1-\zeta)\ell(\ell+1)=C_{\mathrm{LV}}$ and its critical point $\ell_{\mathrm c}$; (e) constrained oscillator frequency $\Omega_{\ell}(a)$; (f)  degree-1 Heun states $\mathcal R_{1}(\xi)$, with $\xi=\sqrt{\widetilde{\Omega}}\,x$, $\widetilde{\Omega}=\Omega/\sqrt{1-\zeta}$ and $\mathcal H_{1}(y)=1+\mathcal{B}_{1}y$.}
\label{fig:2}
\end{figure*}

This equation is the confluent Heun differential equation \cite{Heun1}, with parameter mappings that satisfy the standard relations
\begin{equation*}
\begin{split}
&P=\beta+\gamma+2-\alpha,\qquad
\mu=\frac{1}{2}
\left(
\alpha-\beta-\gamma+\alpha\beta-\beta\gamma
\right)-\eta,\\
&\nu=\frac{1}{2}
\left(
\alpha+\beta+\gamma+\alpha\gamma+\beta\gamma
\right)+\delta+\eta.
\end{split}
\end{equation*}

Upon establishing the formal correspondence with the canonical Heun structure, the general solution is expressed as
\begin{equation}
\begin{split}
\mathcal{H}(y)
&=
\mathcal{C}_1
\mathcal{H}_{C}(\alpha,\beta,\gamma,\delta,\eta,y)
+
\mathcal{C}_2\sqrt{y}\,
\mathcal{H}_{C}(\alpha,-\beta,\gamma,\delta,\eta,y),\\
y&=-\frac{x^2}{\tilde a^2},
\qquad
\alpha=\tilde a^2\tilde\Omega,
\qquad
\beta=-\frac{1}{2},
\qquad
\gamma=0,\\
\delta&=-\frac{\mathcal{E}\tilde a^2}{4},
\qquad
\eta=
\frac{
\mathcal{E}\tilde a^2-\tilde\ell(\tilde\ell+1)+1
}{4},
\end{split}
\label{2.6}
\end{equation}
where $\mathcal{H}_{C}$ denotes the confluent Heun function.

At this stage, the physical consideration that the wave function should not die out at the wormhole throat ($x=0=y$) is important. Since the effective potential (\ref{eff}) does not provide an infinite repulsive core at $x=0$, no asymptotic behavior should be sought at $x=0=y$. Moreover, the asymptotic behavior of both independent solutions should be sought at $x=\pm\infty\Rightarrow y\sim\infty$, which is readily provided by the exponential term in $\mathcal{R}(x)=e^{-{\tilde\Omega x^2}/{2}}\mathcal{H}(x)$ (see also (\ref{2.61}) below). This immediately suggests that the second independent solution should therefore be discarded by setting $\mathcal{C}_2=0$. Consequently, the physically acceptable solution assumes the form
\begin{equation}
\mathcal{R}(y)
=
\mathcal{C}\,
e^{{\tilde a^2\tilde\Omega y}/{2}}\,
\mathcal{H}_{C}(\alpha,\beta,\gamma,\delta,\eta,y),
\label{2.61}
\end{equation}
with the series representation \cite{Heun1}
\begin{equation}
\mathcal{H}_{C}(\alpha,\beta,\gamma,\delta,\eta,y)
=
\sum_{j=0}^{\infty}\mathcal{B}_j y^j,
\end{equation}
where
\begin{equation}
\mathcal{B}_j
=
\frac{C_j}{j(j+P-1)-\mu}.
\end{equation}

Substituting into Eq.~(\ref{2.5}) yields
\begin{equation}
C_1
=
\frac{P-\mu}{\beta+1}C_0,
\qquad
C_0=-\mu\mathcal{B}_0
\Rightarrow
\mathcal{B}_1=-\frac{\mu}{\beta+1},
\qquad
\mathcal{B}_0=1,
\label{2.62}
\end{equation}
and a three-term recurrence relation
\begin{equation}
\frac{
C_{j+2}\left[(j+2)(j+\beta+2)\right]
}{
(j+2)(j+P+1)-\mu
}
=
C_{j+1}
+
\frac{
C_j\left[(\mu+\nu)+\alpha j\right]
}{
j(j+P-1)-\mu
},
\qquad
j\geq0.
\label{2.7}
\end{equation}

Finiteness and square integrability of the quantum wave function require that the infinite power series be truncated to a finite polynomial of some degree. One should observe that the $(n+1)^{\mathrm{th}}$-order recursion relation reads
\begin{equation}
\frac{
C_{n+3}\left[(n+3)(n+\beta+3)\right]
}{
(n+3)(n+P+2)-\mu
}
=
C_{n+2}
+
\frac{
C_{n+1}\left[(\mu+\nu)+\alpha(n+1)\right]
}{
(n+1)(n+P)-\mu
}.
\label{2.71}
\end{equation}

Hence, truncation of the power series to a polynomial of degree $\tilde n=n+1\geq1$ enforces the conditions
\[
C_{n+1}\neq0,
\qquad
C_{n+2}=0=C_{n+3}=C_{n+4}=\cdots.
\]
Consequently, $C_{n+1}\neq0$ implies
\begin{equation}
\mu+\nu=-\alpha(n+1),
\qquad
\delta=-\alpha
\left[
(n+2)+\frac{1}{2}(\beta+\gamma)
\right],
\label{2.8}
\end{equation}
which is in complete agreement with the standard structure of confluent Heun polynomial constraints \cite{Heun1}. Hence, by substituting $\delta$ of Eqs.~(\ref{2.6}) in (\ref{2.8}), the corresponding energy spectrum of the $\mathcal{PT}$-symmetric scalar bosonic field in the LV wormhole background is obtained as
\begin{equation}
\mathcal{E}
=
4\tilde{\Omega}
\left(
n+\frac{7}{4}
\right)
\Rightarrow
E
=
\pm
\left[
4\frac{\Omega}{\sqrt{1-\zeta}}
\left(
n+\frac{7}{4}
\right)
+m_\circ^2
\right]^{1/2},
\qquad
n\geq0.
\label{2.801}
\end{equation}

Moreover, the condition
\begin{equation}
C_{n+2}
=
0
=
\mathcal{B}_{n+2}
\left[
(n+2)(n+P+1)-\mu
\right],
\label{2.802}
\end{equation}
suggests that
\begin{equation}
\mu=(n+2)(n+P+1),
\label{2.81}
\end{equation}
is, in fact, a co-factorized root (consequently, gives $\eta$ in (\ref{2.6}) as a root) for $C_{n+2}=0$ and provides a parametric correlation that involves $(\Omega,a,\ell)$ and the LV parameter $\zeta$. This condition not only secures the truncation of the power series to a polynomial of degree $(n+1)$, but also identifies the conditional exact solvability \cite{Levai} of the confluent Heun equation. However, using $\mu$ from (\ref{2.5}) in (\ref{2.81}) leads to
\begin{equation}
\Omega(n,\ell,\zeta)
\equiv
\Omega
=
\frac{
(2n+5)(2n+4)-(1-\zeta)\ell(\ell+1)
}{
2a^{2}\sqrt{1-\zeta}
}
>0.
\label{2.11}
\end{equation}

This relation reveals a nontrivial interdependence between the oscillator frequency $\Omega$, the wormhole throat scale $a$, the Lorentz-violation parameter $\zeta$, and the quantum numbers $(n,\ell)$. In particular, it suggests an upper bound on the allowed angular momentum quantum number,
\begin{equation}
\ell_{\max}
=
-\frac{1}{2}
+
\sqrt{
\frac{1}{4}
+
\frac{(2n+5)(2n+4)}{1-\zeta}
}
\geq0,
\label{2.111}
\end{equation}
thereby demonstrating that the spectrum is not freely adjustable but instead confined to a restricted parameter manifold dictated by the Heun polynomial structure. Consequently, the spectral problem admits exact solutions only under these stringent algebraic constraints, exemplifying the paradigm of conditional exact solvability \cite{Levai}.

Nevertheless, one should also be aware of the critical $n^{\mathrm{th}}$-order recursion relation that, with $C_{n+2}=0$ in (\ref{2.7}), reads
\begin{equation}
C_{n+1}
=
\frac{\alpha C_n}{n(n+P-1)-\mu}
=
-\frac{\alpha C_n}{4n+2P+2}.
\label{2.82}
\end{equation}
This equation will be responsible for introducing parametric correlations as exemplified in the sequel.

\subsection{Degree-1 Heun polynomial correspondence: conditional exact solvability}

At this point, we consider the most nontrivial case $n=0$ (i.e., a polynomial of degree $1$). Next, we use (\ref{2.82}) and (\ref{2.62}), along with $\mu$ in (\ref{2.81}) and $P=3/2-\alpha$, to obtain
\begin{equation}
\frac{\alpha}{2P+2}
=
\frac{P+2}{\beta+1},
\quad
\beta=-\frac{1}{2}
\Rightarrow
\alpha=\frac{25\mp\sqrt{65}}{8}.
\label{2.812}
\end{equation}

In what follows, we adopt \(\alpha=(25-\sqrt{65})/8\) and proceed by comparing this result with that in (\ref{2.11}) to imply that
\begin{equation}
\frac{25-\sqrt{65}}{8}
=
\frac{20-\tilde\ell(\tilde\ell+1)}{2}
\Rightarrow
(1-\zeta)\ell(\ell+1)
=
\frac{55+\sqrt{65}}{4}.
\label{2.83}
\end{equation}

Obviously, this gives a correlation between the angular momentum quantum number $\ell$ and the LV parameter $\zeta$. Moreover, it suggests that $\tilde\ell\ne0$ and all states with $\ell=0$ are therefore dismissed from the spectrum. This is closely related to the conditional exact solvability of the confluent Heun series/polynomial construction used here.

Under such restrictions, and for $n=0$ in (\ref{2.801}), we obtain
\begin{equation}
E
=
\pm
\left[
7\frac{\Omega}{\sqrt{1-\zeta}}+m_\circ^2
\right]^{1/2}
=
\pm
\left[
7\frac{\Omega\sqrt{4\ell(\ell+1)}}
{\sqrt{55+\sqrt{65}}}
+m_\circ^2
\right]^{1/2}.
\label{2.91}
\end{equation}
where $\Omega$ is given by (\ref{2.11}) as
\begin{equation}
\Omega
=
\frac{
80-(55+\sqrt{65})
}{
8a^{2}\sqrt{1-\zeta}
}
>0.
\label{2.92}
\end{equation}

Within our series-to-polynomial construction considered here, the resulting eigenvalues remain real and symmetric under $E\rightarrow -E$, reflecting the intrinsic relativistic charge-conjugation structure of the KG framework, preserved despite the presence of curvature and LV deformations. In Table~\ref{tab:allowed}, we give a sample of the allowed angular momentum quantum numbers $\ell$ and the corresponding LV parameter $\zeta$ that simultaneously satisfy (\ref{2.82}) and (\ref{2.11}). We clearly observe that $\ell\geq4$ are the allowed values for $\ell$, while states with $\ell<4$ are not allowed. Consequently, the spectral problem admits exact solutions only under these stringent algebraic constraints, exemplifying the paradigm of conditional exact solvability \cite{Levai}.
\begin{table}[t] \centering \setlength{\tabcolsep}{28pt} \caption{ \fontsize{7.6}{8.6}\selectfont Representative allowed integer angular-momentum quantum numbers $\ell$ and the corresponding LV parameters $\zeta_\ell$, determined by $(1-\zeta_\ell)\ell(\ell+1)=C_{\mathrm{LV}}$, where $C_{\mathrm{LV}}=(55+\sqrt{65})/4$. The admissibility condition $0\leq\zeta_\ell<1$ requires $\ell\geq4$.} \label{tab:allowed} \begin{tabular}{cc} \toprule $\ell$ & $\zeta_\ell$ \\ \midrule 4 & 0.211721778 \\ 5 & 0.474481185 \\ 6 & 0.624629418 \\ 7 & 0.718472064 \\ 10 & 0.856676687 \\ 15 & 0.934310148 \\ 20 & 0.962462942 \\ 25 & 0.975745285 \\ 50 & 0.993817426 \\ \bottomrule \end{tabular} \end{table}

\setlength{\parindent}{0pt}
\vspace{1pt}

Fig.~\ref{fig:2} shows that the exact spectrum is confined to a conditional-exact solvability manifold rather than spanning an unconstrained parameter space. The particle--antiparticle symmetry $E^{-}=-E^{+}$ is preserved throughout, while the allowed integer angular momenta are correlated with the LV parameter through $(1-\zeta)\ell(\ell+1)=C_{\mathrm{LV}}$, thus excluding $\ell<4$ from the degree-1 sector. The constrained frequency $\Omega_{\ell}\propto a^{-2}$ shows that increasing the wormhole throat weakens the effective oscillator confinement and drives $E^{+}$ towards the rest-energy threshold $m_{\circ}$. In contrast, increasing $\zeta$ requires progressively larger admissible $\ell$, while $\widetilde{\Omega}=\Omega/\sqrt{1-\zeta}$ controls the corresponding localization scale of the Heun states. The resulting wave functions remain normalizable and localized, with their profiles varying systematically across the admissible $\ell$ sectors. Thus, Lorentz violation, wormhole geometry, angular momentum, and oscillator confinement are not independent parameters in the exact sector; instead, they are algebraically correlated by the Heun polynomial conditions, yielding a restricted but analytically tractable family of relativistic bound states.

Nevertheless, it is unavoidable to mention that above we have adopted \(\alpha=(25-\sqrt{65})/8\) and obtained the parametric constraint \((1-\zeta)\ell(\ell+1)={(55+\sqrt{65})}/{4}\), which allowed angular momentum quantum number \(\ell\geq4\). On the other hand, one may wish to work with \(\alpha=(25+\sqrt{65})/8\) and obtain a parametric constraint  \((1-\zeta)\ell(\ell+1)={(55-\sqrt{65})}/{4}\), which would allows \(\ell\geq3\). In both cases, however, the overall physical behavior remains the same.

\bigskip
\section{Summary and Discussion}
\label{sec:4}

\setlength{\parindent}{0pt}

We have studied a non-Hermitian $\mathcal{PT}$-symmetric KG oscillator in a LV traversable wormhole background. The constant lapse $\mathcal{A}(x)=1$ and the LV-dependent areal radius in Eq.~\eqref{2.01} provide a regular two-sided geometry with a finite throat at $x=0$. As shown in Fig.~\ref{fig:introduction}, increasing $\zeta$ leaves the throat radius unchanged while modifying the radial geometry and enhancing the effective oscillator scale $\widetilde{\Omega}=\Omega/\sqrt{1-\zeta}$.

\setlength{\parindent}{0pt}
\vspace{1pt}

The resulting radial dynamics is governed by the effective potential in Eq.~\eqref{eff}, which is finite at the throat and grows as $\widetilde{\Omega}^{2}x^2$ for large $|x|$. Hence, the wormhole geometry supports regular localized modes on the complete radial line. The dependence of the potential on the LV parameter, angular momentum, and oscillator frequency is displayed in Fig.~\ref{fig:1}. In contrast to the ordinary flat-space radial problem, the throat is an interior regular point, allowing the radial modes to extend continuously between the two asymptotic regions.

\setlength{\parindent}{0pt}
\vspace{1pt}

The spectral equation is reduced to the confluent-Heun form in Eq.~\eqref{2.5}, with the corresponding Heun representation given by Eq.~\eqref{2.6}. Polynomial truncation through the recurrence conditions in Eqs.~\eqref{2.7} and \eqref{2.71} leads to conditional exact solvability: the model parameters cannot be chosen independently but must satisfy algebraic correlations. For the degree-1 sector, these conditions correlate the angular momentum with the LV parameter, yielding the admissible relation used to construct the states reported in Table~\ref{tab:allowed}. The requirement $0\leq\zeta<1$ then restricts this exact sector to integer $\ell\geq4$; this restriction applies only to the degree-1 polynomial sector and not to the underlying radial equation.

\setlength{\parindent}{0pt}
\vspace{1pt}

The resulting spectrum is real and occurs in particle--antiparticle pairs satisfying $E^-=-E^+$. The conditional spectrum, parameter correlations, and degree-1 wave functions are displayed in Fig.~\ref{fig:2}. In particular, the exact solutions demonstrate that Lorentz violation modifies the oscillator confinement and the admissible angular-momentum sectors simultaneously, while the throat radius controls the geometric scale of the bound states. The reality of the energies is obtained within the algebraically constrained $\mathcal{PT}$-symmetric sector rather than for arbitrary values of the model parameters.

\setlength{\parindent}{0pt}
\vspace{1pt}

The analysis establishes a conditionally exactly solvable relativistic bound-state problem in which LV geometry, wormhole topology, and non-Hermitian $\mathcal{PT}$-symmetric dynamics are intrinsically coupled. The combination of the regular effective potential in Eq.~\eqref{eff}, the confluent-Heun reduction in Eq.~\eqref{2.5}, and the polynomial conditions of Sec.~\ref{sec:3} provides an analytically controlled framework for localized KG modes in LV wormhole spacetimes.


\end{document}